\documentclass[conference]{IEEEtran}
\IEEEoverridecommandlockouts
\usepackage{lipsum}

\usepackage{cite}
\usepackage{amsmath,amssymb,amsfonts}
\usepackage{algorithmic}
\usepackage{graphicx}
\usepackage{textcomp}
\usepackage{xcolor}

\usepackage[compact]{titlesec}
\titlespacing{\section}{0pt}{2ex}{1ex}
\titlespacing{\subsection}{0pt}{2ex}{1ex}
\titlespacing{\subsubsection}{0pt}{2ex}{1ex}

\def\BibTeX{{\rm B\kern-.05em{\sc i\kern-.025em b}\kern-.08em
    T\kern-.1667em\lower.7ex\hbox{E}\kern-.125emX}}

\graphicspath{ {./images/} }

\begin{document}

\title{Exploring the Effects of Multicast Communication on DDS Performance\\
\thanks{This work was partially supported by the project Aggregated Quality Assurance for Systems (AQUAS H2020-EU.2.1.1.7 ID: 737475) \cite{aquas_project}.}
}

\author{\IEEEauthorblockN{Kaleem Peeroo}
\IEEEauthorblockA{\textit{Department of Computer Science} \\
\textit{City, University of London}\\
Kaleem.Peeroo@city.ac.uk}
\and
\IEEEauthorblockN{Peter Popov}
\IEEEauthorblockA{\textit{Department of Computer Science} \\
\textit{City, University of London}\\
P.T.Popov@city.ac.uk}
\and
\IEEEauthorblockN{Vladimir Stankovic}
\IEEEauthorblockA{\textit{Department of Computer Science} \\
\textit{City, University of London}\\
Vladimir.Stankovic.1@city.ac.uk}
}

\maketitle
\begin{abstract}
The Data Distribution Service (DDS) \cite{dds_spec_v1.4} is an Object Management Group (OMG) \cite{omg} standard for high-performance and real-time systems. DDS is a data-centric middleware based on the publish-subscribe communication pattern and is used in many mission-critical, or even safety-critical, systems such as air traffic control and robot operating system (ROS2) \cite{ros2}.

This research aims at identifying how the usage of multicast affects the performance of DDS communication for varying numbers of participants (publishers and subscribers). The results show that DDS configured for multicast communication can exhibit worse performance under a high load (a greater number of participants) than DDS configured for unicast communication. This counter-intuitive result reinforces the need for researchers and practitioners to be clear about the details of how multicast communication operates on the network.
\end{abstract}


\section{Introduction}
DDS is a middleware standard published by the OMG for high-performance, real-time, and scalable systems. This standard follows a publish-subscribe communication pattern and is used in various applications that need real-time communication, e.g. air-traffic control, robotics, autonomous vehicles, etc. In DDS, participants - \textit{publishers} and \textit{subscribers} - communicate with each other via one or more pieces of information referred to as \textit{topics}. The DDS specification defines 22 Quality-of-Service (QoS) parameters to customise the communication, e.g. \textit{Reliability}, which controls whether publishers await an acknowledgement from subscribers before publishing the following data sample  - \textit{Reliable} communication (or conversely \textit{"best effort"} communication). There are also non-QoS parameters that customise the communication, e.g. the use of unicast or multicast communication. \textit{Unicast} communication defines that a host will only send data to another single host at any time while \textit{multicast} establishes the use of one-to-many communication whereby a single host can send to multiple hosts in a single communication.

\section{Problem Definition}
Despite not being defined within the DDS specification, the use of unicast or multicast still plays a vital role in how the communication of messages within DDS takes place. DDS users who have limited knowledge of the details of multicast implementation may therefore assume that multicast transport should, theoretically, outperform that of unicast for greater numbers of participants since the data are being sent in one communication rather than multiple. This, however, does not necessarily hold in practice - we explore this counter-intuitive phenomenon in the paper.

Multicast Routing Protocols (MRP) can vary between different setups and this can affect the performance. Essentially, MRPs create a shortest path tree (SPT) from the sender to the receiver group. There are two types of SPTs. The performance varies depending on how the network is structured and what type of SPTs are created.

This research focuses on the effects of multicast on DDS performance under varying
numbers of participants. With the results produced from this research, we can determine that, although it is intuitive to use multicast for a large number of participants, one should have a clear understanding of i) the network topology and ii) the way multicast works to determine if using multicast is beneficial.

\begin{figure*}[!t]
    \centering
    \includegraphics[scale=0.5]{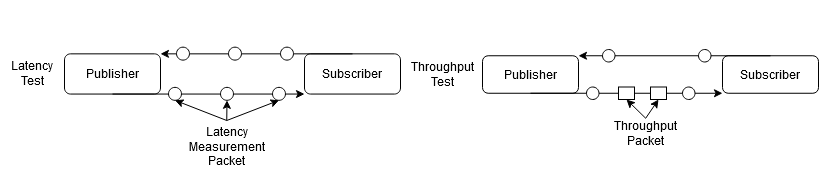}
    \caption{Latency vs throughput tests.}
    \label{fig:latency_vs_throuhpugt_test}
\end{figure*}

\begin{figure*}
    \includegraphics[width=\textwidth]{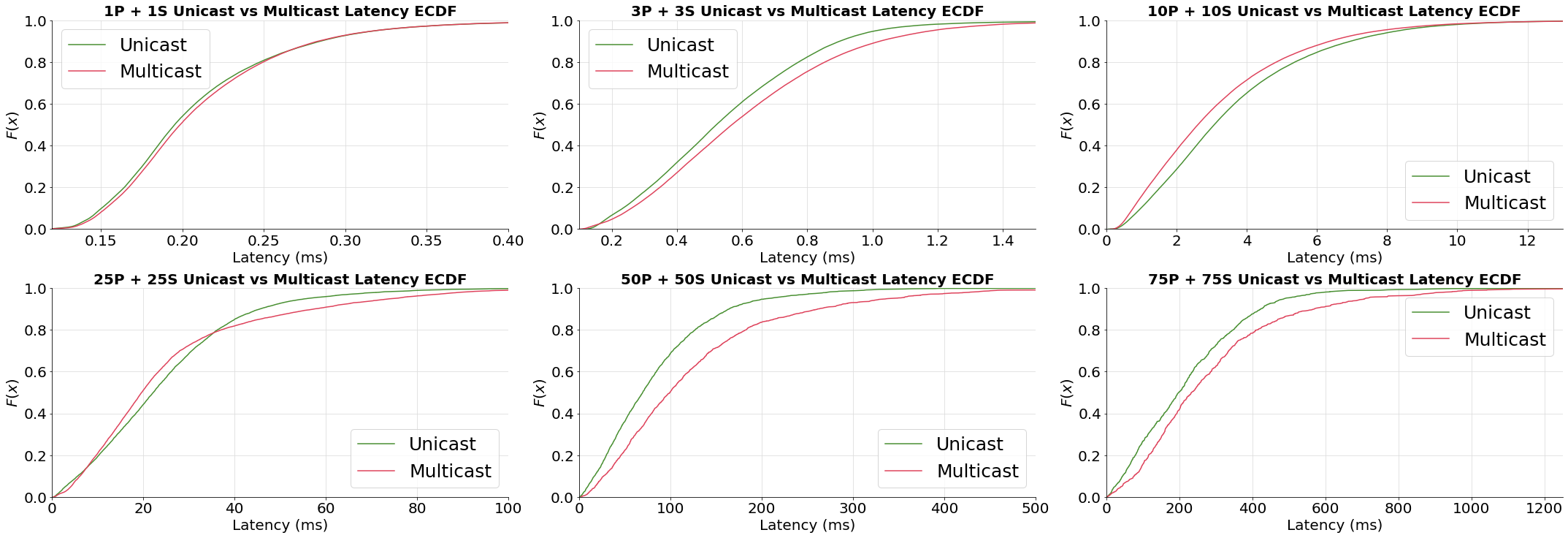}
    \caption{Latency ECDFs for varying participant numbers.}
    \label{fig:latency_cdf_plots}
\end{figure*}

\section{Experimental Setup}
The experimental setup used to carry out these performance tests include 4 virtual machines running CentOS 7.9-2009 with 8GB of RAM and 500GB of hard disk space. For the performance tests, a tool called PerfTest \cite{rti_perftest} provided by Real-Time Innovations (RTI) \cite{rti} was utilised which allows the use of two types of tests: \textit{Latency} or \textit{Throughput}. We use the latter type of test though both types of tests measure latency and throughput. The Throughput test starts with a particular publisher (Publisher 0) sending a latency measurement packet and awaiting its response before sending a user-defined number of throughput packets to the subscriber/s imitating the real-world usage of a DDS system. Once this amount of throughput packets has been transmitted, the publisher will repeat the process (by sending another latency measurement packet). The 2-way latency is measured on Publisher 0, and the throughput is measured by each subscriber recording its received samples over time. Therefore, within our measurements, we have aggregated the throughput for all subscribers per time unit to fairly compare the throughput measurements per test. Figure \ref{fig:latency_vs_throuhpugt_test} provides a brief overview of how the packets are transmitted for each type of test. Within the experiments, values have been varied for the data length, test duration, reliability, participant amount, and communication type. Table \ref{table:experiment_configuration} shows the settings used per participant. We have allocated the participants equally per virtual machine to ensure a fair communication load.

The tests of this paper are a part of our comprehensive campaign for experimental evaluation of DDS performance, which has been inspired by an air traffic control scenario in the AQUAS project \cite{aquas_project} where various numbers of drones were experimented with, ranging from single to double digits - a real-life application. These specific tests produced strange results which have resulted in the creation of this paper.

\begin{table}[h!]
    \centering
    \caption{Configuration settings of the experiments}
    \label{table:experiment_configuration}
    \begin{tabular}{|c|c|}
        \hline
        \textbf{Setting} & \textbf{Value} \\
        \hline
        Data Length & 100 Bytes \\
        Test Duration & 6 Hours \\
        Latency Count & 1000 \\
        Reliability & \verb|reliable| \\
        Number of Publishers & 1, 3, 10, 25, 50, 75 \\
        Number of Subscribers & 1, 3, 10, 25, 50, 75 \\
        \hline
    \end{tabular}
\end{table}

The ``Latency Count'' setting defines how many throughput packets are transmitted between each latency measurement packet. The ``Reliability'' setting was set to \verb|reliable| as this is what is expected to be used in critical systems where the data transmission must be confirmed to function.

\begin{figure*}
    \centering
    \includegraphics[width=\textwidth]{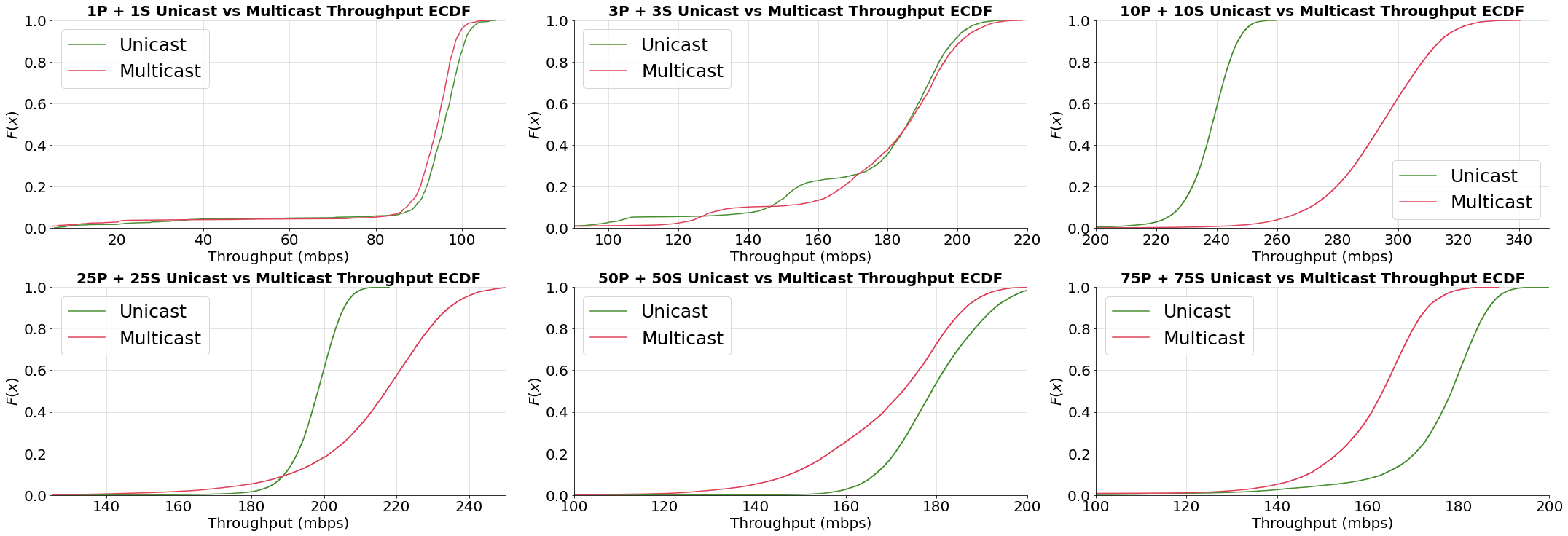}
    \caption{Throughput ECDFs for varying participant numbers.}
    \label{fig:throughput_cdf_plots}
\end{figure*}

\section{Experimental Results}
Figure \ref{fig:latency_cdf_plots} displays the experimental cumulative distribution functions (ECDF) for the latency measurements of each test and Figure \ref{fig:throughput_cdf_plots} shows the ECDF of the throughput measurements. If one were to examine the top left ECDF in Figure \ref{fig:latency_cdf_plots}, the point (0.25, 0.8) can be interpreted as ``80\% of all latency measurements had a value of 0.25ms or less''. Throughout this paper the following abbreviations are used: ``ms'' - milliseconds, ``$\mu$s'' - microseconds, and ``mbps'' - megabits per second.

\subsection{Latency Results}

In the tests with 1 publisher and 1 subscriber the unicast communication produced lower latency values than the multicast communication even though the distribution functions were almost identical. The unicast and multicast latencies both had a mean of 0.22ms. In the tests with 3 publishers and 3 subscribers, the unicast communication values had a mean of 0.56ms whilst the multicast communication values had a mean of 0.62ms - an 11\% increase. For the tests with 10 publishers and 10 subscribers, the ranking was reversed: the mean unicast latency was 3.6ms whilst the mean multicast latency was 3.2ms - an 11\% decrease. Whilst the results from the previous tests clearly showed stochastic ordering (there is no crossover point between the curves on the ECDFs) between the two communication types, the test with 25 publishers and 25 subscribers had none. The unicast communication contained more values between 0ms to 10ms as well as values above 35ms whilst the multicast communication results contained more values between 10ms to 35ms. The next test, including 50 publishers and 50 subscribers, continued the pattern of showing stochastic dominance for the unicast communication where the distribution function shows that the multicast communication consistently produced higher latencies. The unicast communication had a latency mean of 8.3ms whilst the multicast communication had a latency mean of 12.5ms - a 51\% increase. In the final test with 75 publishers and 75 subscribers, the distribution function shows consistently lower latency values for the unicast communication (with a mean of 312ms) compared to multicast communication (having a mean of 427ms - a 37\% increase).

In summary, the unicast communications produced better latencies for the tests involving 1 publisher and 1 subscriber, 3 publishers and 3 subscribers, 50 publishers and 50 subscribers, and 75 publishers and 75 subscribers. Of the 6 tests, 4 concluded with the unicast setting producing lower latency measurements. Only 1 test (10 publishers and 10 subscribers) produced better multicast latency results. The test with 25 publishers and 25 subscribers is the only test that produced results where neither setting was consistently dominant. 

\subsection{Throughput Results}
In Figure \ref{fig:throughput_cdf_plots}, the tests involving 1 publisher and 1 subscriber show that roughly 10\% of the throughput measurements were almost identical. For throughput measurements above 85mbps, the unicast communication produced higher throughput values consistently. In the tests with 3 publishers and 3 subscribers, neither communication setting was dominant. The unicast communication had a higher percentage of values up to 125mbps whilst the multicast communication produced a higher percentage of values between 125mbps and 145mbps. The unicast communication measurements contained more values between 145mbps to 170mbps while the multicast communication measurements contained more values between 170mbps to 185mbps, and for values greater than 185mbps, unicast had a higher percentage. In the tests with 10 publishers and 10 subscribers, the ECDFs show the multicast communication producing a larger throughput. The mean throughput value for the unicast communication was 237mbps whilst for the multicast communication it was 292mbps - a 23\% increase. The tests with 25 publishers and 25 subscribers also did not show a clear difference: the multicast communication produced a higher percentage of values up to 190mbps before this observation reverses and the unicast communication produces a much higher percentage of values up to 220mbps. The unicast throughput had a maximum value of 217 mbps whilst the multicast throughpout had a maximum value of 260 mbps. The results produced from the ``50 publishers / 50 subscribers'' tests demonstrated the multicast communication consistently producing lower throughputs than the unicast communication. The mean of the unicast measurements was 179mbps compared to the multicast mean of 169mbps - a 6\% decrease. The tests with 75 publishers and 75 subscribers show a similar pattern in which the multicast results produced consistently lower throughput measurements than the unicast communication. The unicast throughput mean was 175mbps while the multicast mean was 160mbps - a 9\% decrease.

In summary, unicast produced better throughput values for tests involving part of 1 publisher and 1 subscriber, 50 publishers and 50 subscribers and 75 publishers and 75 subscribers. Meanwhile, multicast produced better throughput values for 10 publishers and 10 subscribers and most of 25 publishers and 25 subscribers. Out of the 6 tests, the unicast communication produced more performant measurements for 3 of the tests while the multicast communication produced more performant throughput values for 2 of the tests. The test with 3 publishers and 3 subscribers was the only test where no consistent ranking was observed.

\section{Discussion}
Theoretically, when a large number of participants uses multicast communication, all of the recipients' data are sent using one communication rather than multiple (as in the case of unicast) - this should result in better performance of multicast with a large number of participants. Our results demonstrate the opposite. A summary showing the most performant communication is shown in Table \ref{table:unicast_vs_multicast_results}. After investigating the cause of this, we conjecture that performance gained when using multicast communication with DDS largely depends on the network setup. Specifically, which multicast routing protocol is being used. To confirm this, we plan to experiment further and vary the network structure to mirror real-life applications whilst investigating the effects of multicast communication. The main point from this work is that one must have a thorough understanding of how multicast communication is implemented on one's (experimental) setup before using it for DDS deployment. One should not assume that using multicast via a DDS implementation with large numbers of participants will necessarily result in better performance. 

\begin{table}[h!]
    \centering
    \caption{Most performant communication type per test}
    \label{table:unicast_vs_multicast_results}
    \begin{tabular}{|c|c|c|}
        \hline
         & \textbf{Latency} & \textbf{Throughput} \\
        \hline
        1P + 1S & \verb|unicast| & \verb|unicast| \\
        3P + 3S & \verb|unicast| & mixed \\
        10P + 10S & \verb|multicast| & \verb|multicast| \\
        25P + 25S & mixed & \verb|multicast| \\
        50P + 50S & \verb|unicast| & \verb|unicast| \\
        75P + 75S & \verb|unicast| & \verb|unicast| \\
        \hline
    \end{tabular}
\end{table}

\section{Related Work}
Whilst many papers have evaluated DDS performance under varying settings, only a few  have focused on multicast.

The research demonstrated in \cite{a_dds_based_energy_management_framework_for_small_microgrid_operation_and_control} experiments with a reliability setting of ``Best Effort'' when using 8 publishers and 6 subscribers. The results show that the average latency of unicast is 243$\mu$s while for multicast it is 270$\mu$s. Another paper, \cite{a_study_publish_subscribe_middleware_under_different_iot_traffic_conditions} investigated varying data lengths and in the case of 1 publisher and 1 subscriber, the average unicast latency was greater than the average multicast latency for a data length of 128 bytes (close to the 100 bytes used in our experiments). In terms of the throughput results, the unicast and multicast communication were almost identical with the average unicast throughput being slightly greater. In the case with 1 publisher and 7 subscribers, the average multicast throughput was roughly four times larger than the average unicast throughput and the average unicast latency was significantly larger.

In \cite{evaluating_a_prototype_approach_to_validating_a_dds_based_system_architecture_for_automated_manufacturing_environments}, the experiments focused on 1 publisher and a variation of 4 and 12 subscribers (along with different data lengths). In the case with 4 subscribers, the average unicast throughput was almost identical to the average multicast throughput with the latter being slightly lower. The difference between the throughput measurements was much more significant in the test with 12 subscribers: the multicast throughput measurements were greater. In \cite{dds_based_interoperability_framework_for_smart_grid_testbed_infrastructure}, the authors experimented with 4 publishers and 3 subscribers under varying data lengths. When 128-byte data packets were used, the average unicast latency was $\sim$ 300$\mu$s whilst the average multicast latency $\sim$ 350$\mu$s.

All of the literature mentioned within this section have quite different values from the measurements attained within this research. We suspect this may be due to the virtualisation used in our experiments.

\section{Conclusion}
This paper has investigated the effects of multicast communication on a DDS implementation using RTI Perftest whilst varying the number of participants. The results demonstrate a counter-intuitive phenomenon: when a large number of participants is used, the use of unicast communication outperforms multicast communication.

We conjectured that the multicast performance fundamentally depends on the network settings, and especially which multicast routing protocol is used. If true, the inferior performance of multicast, for some test types, is thus not due to the DDS protocol nor the specific DDS implementation used. The results of this paper, therefore, motivate our further research into understanding the details of how multicast communication works with DDS and how changing the various aspects of this communication may affect the performance. With this in mind, we intend  to run further experiments with different network settings including the multicast protocol as well as varying other DDS-related settings, e.g. using different data lengths with an unequal number of participants to investigate the effect of multicast communication under different writing/reading loads and evaluating multicast communication on other DDS implementations. We also plan to explore what effect(s) virtualisation has on the performance of multicast communication.

% The results of this paper, therefore, motivate our further research on understanding the intricacies of the network structure and the multicast routing protocol being used. Also, our future work includes further varying experimental settings, e.g. using different data lengths and using an unequal number of publishers and subscribers to investigate the impact of multicast under different writing/reading loads. We also plan on evaluating the effects of multicast on the performance of other DDS implementations. Furthermore, we plan on experimenting with the use of different multicast routing protocols, and their effect on DDS performance. It would also be interesting to identify what effect virtualisation has on the use of unicast and multicast especially since virtualisation may distort measurements under a high load.

\bibliographystyle{IEEEtran}
\bibliography{references}

\end{document}